\documentclass[preprint,aps,showpacs]{revtex4}
\usepackage{latexsym}
\begin{document}
%
\title{GENERALIZED DISCRETIZATION OF THE KARDAR-PARISI-ZHANG EQUATION}
\author{R. C. Buceta}

\affiliation{Departamento de F\'{\i}sica, Facultad de Ciencias
Exactas y Naturales, Universidad Nacional de Mar del Plata\\Funes
3350, B7602AYL Mar del Plata, Argentina}


\begin{abstract}
We introduce the generalized spatial discretization of the
Kardar-Parisi-Zhang (KPZ) equation in 1+1 dimensions. We solve
exactly the steady state probability density function for the
discrete heights of the interface, for any discretization scheme.
We show that the discretization prescription is a consequence of
each particular model. From the ballistic deposition model we
derive the discretization  prescription of the corresponding KPZ
equation.
\end{abstract}

\maketitle

The interface growth models has attracted much attention during
the two last decades due to its widespread application to many
systems \cite{barabasi,halpin,meakin}, such as film growth by
vapor or chemical deposition, bacterial growth, evolution of
forest fire fronts, etc. For such systems, the major effort has
been concentrated in the identification of the scaling regimes and
their classification into universality classes through Monte Carlo
simulation of the discrete models or renormalization-group
analysis of the continuous equations that describe the evolution
of the interfaces in the coarse-grained approximation. This
powerful tool allows a correspondence between the discrete models
and the continuous equations. Another useful method is to derive
continuous evolution equations from the transition rules of the
discrete growth models based on a regularizing scheme and
coarse-graining of the discrete Langevin equations
\cite{vvedensky,costanza,braunstein}. Phenomenological equations,
selected according to symmetry principes and laws conservation,
are often able to reproduce various experimental data. Between
these phenomenological equations the introduced by Kardar, Parisi,
and Zhang (KPZ) has been succesful \cite{kardar} to describe the
properties of  rough interfaces, but is also related to Burgers
equation of turbulence, and to directed polymers in random media
\cite{halpin}. The KPZ equation describes the evolution of the
profile $h(x,t)$ of the interface at position $x$ and time $t$:
\begin{equation}
\frac{\partial h}{\partial t}=\nu \;\frac{\partial^2 h}{\partial
x^2}+ \frac{\lambda}{2} \left(\frac{\partial h}{\partial
x}\right)^2 + \eta(x,t)\;,
\end{equation}
where $\nu$ and $\lambda$ are the diffusion and nonlinear
coefficients, respectively. The Gaussian thermal noise $\eta(x,t)$
has zero mean and covariance
\begin{equation}\label{Eq.continuum-cov}
\langle\eta(x,t)\eta(x',t')\rangle=2\,\epsilon\;\delta(x -
x')\delta (t - t')\;,
\end{equation}
where $\epsilon$ is the noise intensity. Here and elsewhere
$\langle\hspace{.5ex}\rangle$ denotes average over the noise
realizations. The KPZ equation differs from the Edwards-Wilkinson
(EW) equation \cite{edwards}, that describe interfaces growing
under the effect of random deposition and surface tension, in a
nonlinear term due to microscopic lateral growth. The direct
numerical integration is a powerful and simple tool to compute the
exponents that characterize the universality class of a given
continuous equation \cite{amar1,moser,beccaria,dasgupta}. In this
context, the KPZ equation has been integrated using a unusual
discretization method with exact steady state probability density
function \cite{lam}, a pseudospectral discretization method
\cite{giada}, and a least-square error method from experimental
data \cite{giacometti}. Several direct numerical integration
methods has been used in various growing models
\cite{amar1,moser,beccaria,dasgupta,lam} although the theoretical
justification is not clear.

The main goal of this Communication is to introduce the
generalized spatial discretization of the KPZ equation in $1+1$
dimensions. We show that exists the steady state probability
density function of the discrete interface heights compatible with
the continuous ones, after coarse grained approximation,
independent of the discretization scheme.  Finally, we find the
discretization prescription for the KPZ equation derived from the
ballistic model.

\emph{Generalized discretization.} We propose as generic spatial
discretization of the KPZ equation the following Langevin equation
\begin{equation}\label{eq:discrete-kpz}
\frac{dh_i}{dt}=\nu\,L_i + \frac{\lambda}{2}\,N_i^{(\gamma)}+
\eta_i(t)\;,
\end{equation}
where $h_i(t)=h(ia,t)$ is the interface height at the
$i$th-lattice point ($i=1,\dots,N$), $a$ is the horizontal lattice
spacing, and $L=a\,N$ is the lattice size. Periodic boundary
conditions are assumed, {\sl i.e.} $h_0\equiv h_N$. Without loss
of generality we take the horizontal and vertical lattice spacing
to be equal. Introducing in Eq.~(\ref{eq:discrete-kpz}) the
addimensional difference of heights
\begin{equation}\label{eq:difference-heights}
H^{i+\ell}_{i+k}=\frac{1}{a}(h_{i+\ell}-h_{i+k})\;,
\end{equation}
where $\ell,k=-1,0,1$ ($\ell\neq k$), the standard discretized
diffusive term is given by
\begin{equation}\label{eq:Gamma}
L_i=\frac{1}{a}\left(H_i^{i+1}-H^i_{i-1}\right)\;.
\end{equation}
Our discretized nonlinear term (with $0\leq\gamma\leq1$) in
Eq.~(\ref{eq:discrete-kpz}) is defined as
\begin{equation}\label{eq:psi-gamma}
N_i^{(\gamma)}\!=\frac{1}{2(\gamma\!+\!1)}\!\left[(H_i^{i+1})^2\!+\!
2\gamma\,H_i^{i+1}H_{i-1}^i\!+\!(H^i_{i-1})^2\right]\!.
\end{equation}
The noise $\eta_i$ has zero mean and covariance
\begin{equation}\label{eq:cov-discrete}
\langle\eta_i(t) \eta_j(t') \rangle =
\frac{2\,\epsilon}{a}\;\delta_{ij}\;\delta (t - t')\;.\nonumber
\end{equation}
Expanding Eq.~(\ref{eq:Gamma}) and Eq.~(\ref{eq:psi-gamma})
around the $i$th-site, the discretized diffusive term and
the nonlinear term in the $\gamma$--discretization are
\begin{eqnarray}\label{eq:diff}
L_i&=&\left.\frac{\partial^2 h}{\partial
x^2}+\frac{1}{12}\;\frac{\partial^4 h}{\partial x^4}\;
a^2+\mathcal{O}(a^4)\right\rfloor_{x=i a}\;,\\
N_i^{(\gamma)}\!&=&\!\left.\left(\frac{\partial h}{\partial
x}\right)^{\!2}\!+\frac{1}{4}\!\left(\frac{1-\gamma}{1+\gamma}\right)
\!\left(\frac{\partial^2 h}{\partial x^2}\right)^{\!2} a^2
+{\cal{O}}(a^4)\right\rfloor_{x=ia},\nonumber
\end{eqnarray}
respectively. Several discretizations will produce different
results in the roughness eventhough the difference in numerical
accuracy of the height profile are small \cite{lam}. Most
numerical studies are done with the discrete spatial version of
the KPZ equation corresponding to the usual choice $\gamma=1$ in
Eq.~(\ref{eq:psi-gamma}) called standard or post-point
discretization. The nonlinear term $N_i^{(1)}=\frac{1}{4}
(H_{i-1}^{i+1})^2$ only depends on the nearest neighbor sites
height and  minimizes the error in approximating ${\partial^2
h}/{\partial x^2}$ [see Eq.~(\ref{eq:diff})]. Moreover, the choice
$\gamma=0$ in Eq.~(\ref{eq:psi-gamma}), called anti-standard or
pre-point discretization, used here afterwards, corresponds to the
arithmetic mean of the squared slopes around any interface site.
On the other hand, Lam and Shin \cite{lam} introduced the spatial
discretization corresponding to the choice $\gamma=1/2$ in
Eq.~(\ref{eq:psi-gamma}) that enables an elegant analytical
treatment. However, this choice is unusual and is only supported
by the existence of a steady state probability density function
equal to those of the linear case $\lambda=0$. We explain this
special choice and show, through a general calculus of the steady
state solution, that the generalized discretization has an
unambiguous limit in the continuous.

\emph{Steady state density.} The main feature of the generalized
discretization of the KPZ equation is that the corresponding
steady state probability density function
$\widetilde{P}(\mathbf{h})$ of the discrete heights
$\mathbf{h}\!\equiv\!\{h_i\}$ exists. This density function give
rise to the known steady state probability density functional
$\widetilde{P}[h]$ of the continuous interface height $h(x,t)$
related to the KPZ equation \cite{halpin}. Conversely,
$\widetilde{P}(\mathbf{h})$ is not a direct consequence of the
$\widetilde{P}[h]$ as we show  bellow. The probability density
function $P(\mathbf{h},t)$ of the discrete interface evolves
according to the following Fokker-Planck equation
\begin{eqnarray}\label{eq:discrete-FP}
\frac{\partial P}{\partial t}&=&\!-\sum_{i=1}^N\frac{\partial
J_i^{(\gamma)}}{\partial h_i\hspace{1ex}}\,,\\
J_i^{(\gamma)}&=&\left(\nu\,L_i +
\frac{\lambda}{2}\,N_i^{(\gamma)}\right)P-
\frac{\epsilon}{a}\;\frac{\partial P}{\partial h_i}\;,\nonumber
\end{eqnarray}
where $J_i^{(\gamma)}\!(\mathbf{h},t)$ is the probability current
density. Replacing in Eq.~(\ref{eq:discrete-FP}) the general
\begin{equation}\label{eq:discrete-stst-gamma}
\widetilde{P}(\mathbf{h})=\exp\left(-\frac{\nu}{\epsilon}\,
\sum_{i=1}^N\,aU_i^{(\gamma)}\right)\,,
\end{equation}
where $U_i^{(\gamma)}$ is a ``potential'' function to be derived
bellow, we obtain the steady state probability current density function
\begin{equation}\label{eq:zero-stst-flux}
\widetilde{J}_i^{(\gamma)}(\mathbf{h})=\left(\nu\,L_i
+\frac{\lambda}{2}\,N_i^{(\gamma)} +\nu\,\frac{\partial
U_i^{(\gamma)}}{\partial h_i\hspace{1ex}}\right)\widetilde{P}\;.
\end{equation}
Lam and Shin \cite{lam} showed that $\widetilde{P}(\mathbf{h})$
for the linear case ($\lambda\!=0$) is also solution of the
nonlinear one in the midpoint discretization. We explain their
result setting $U_i^{(\gamma)}=N_i^{(0)}$ in
Eq.~(\ref{eq:zero-stst-flux}), {\sl i.e.} the solution
corresponding to $\lambda=0$. It is easy to show that, under
periodic boundary conditions, the current
$\widetilde{J}_i^{(\gamma)}\!=
(\lambda/2)\,N_i^{(\gamma)}\,\widetilde{P}\,$ is conserved only if
$\gamma=1/2$ for $\lambda\neq 0$  ({\sl i.e.}
$\sum_{i=1}^N\partial\widetilde{J}_i^{(1/2)} \!/\partial h_i=0$).
Our goal is to find a steady state solution, independent of the
choice of the discretization and therefore for any nonlinear
coefficient $\lambda$, with constant
$\widetilde{J}_i^{(\gamma)}(\mathbf{h})$. It is easy to show,
using Eq.~(\ref{eq:Gamma}) and Eq.~(\ref{eq:psi-gamma}), that
\begin{eqnarray}\label{eq:replace}
N_i^{(\gamma)}&=&N_i^{(1/2)}+\frac{1}{6}
\left(\frac{1-2\gamma}{1+\gamma}\right){L_i}^2\;a^2\nonumber\;,\\
\frac{dL_i}{d h_i}&=&-\frac{2}{\hspace{.5ex}a^2}\;,\\
\frac{dN_i^{(\gamma)}}{d h_i\hspace{1ex}}
&=&-\left(\frac{1-\gamma}{1+\gamma}\right)L_i
\nonumber\;.
\end{eqnarray}
Using Eqs.~(\ref{eq:replace}) in Eq.~(\ref{eq:zero-stst-flux}) we obtain that
\begin{equation}\label{eq:exp-discrete-stst2}
U_i^{(\gamma)}=N_i^{(0)}+\sigma\,{L_i}^3\;,
\end{equation}
where
\[
\sigma=\frac{\lambda}{72\;\nu}\left(\frac{1-2\gamma}{1+\gamma}\right)\;a^4\;.\hspace{4ex}
\]
Eq.~(\ref{eq:exp-discrete-stst2}) gives a steady state solution of
the Fokker-Planck equation [Eq.~(\ref{eq:discrete-stst-gamma})]
with conserved current
\[
\widetilde{J}_i^{(\gamma)}\equiv \widetilde{J}_i^{(1/2)} =
\frac{\lambda}{2}\,N_i^{(1/2)}\,\widetilde{P}\;.
\]
A dimensional analysis shows that $\epsilon/\nu$ and $\nu/\lambda$
are proportional to $a$, and therefore $\sigma$ is proportional to
$a^3$. Besides, note from Eq.~(\ref{eq:diff}) that the errors of
$N_i^{(\gamma)}$ and $L_i$ in approximating $(\partial h/\partial
x)^2$ and $\partial^2h/\partial x^2$, respectively, are at most
proportional to $a^2$. Thus, we conclude that the error of
$U_i^{(\gamma)}$ in approximating $(\partial h/\partial x)^2$ is
proportional to $a^2$ [see Eq.~(\ref{eq:exp-discrete-stst2})]. The
midpoint solution $U_i^{(1/2)}=N_i^{(0)}$ is symmetric under the
interchange $H_i^{i+1}\leftrightarrow H^i_{i-1}$, but the term
$\sigma L_i^3$ breaks weakly this symmetry in
Eq.~(\ref{eq:exp-discrete-stst2}). Notice that the limits
$\lambda=0$ (EW equation), $\gamma=1/2$ (midpoint discretization),
and $a=0$ (coarse grained approximation) in
Eq.~(\ref{eq:exp-discrete-stst2}) are equivalent between them.

If we denote by $P([h],t)$ the probability density functional
of the interface position function $h(x,t)$ in $1+1$ dimensions,
the corresponding Fokker-Planck equation is
\begin{eqnarray*}
\frac{\partial P([h],t)}{\partial t}=&-&\int_0^L
dx\;\frac{\delta\hspace{1ex}}{\delta h}
\left\{\left[\nu\frac{\partial^2h}{\partial x^2}
+\frac{\lambda}{2}\left(\frac{\partial h}{\partial
x}\right)^2\right]P\right\}\\&+&\epsilon\;\int_0^L dx\,
\frac{\delta^2P}{\delta h^2}\;,
\end{eqnarray*}
from which we can obtain the well-known steady state solution
\[
\widetilde{P}[h]=\exp\left[-\frac{\nu}{\epsilon}\,\int_0^L
dx\left(\frac{\partial h}{\partial x}\right)^{\!2}\right]\;.
\]
Incidentally, this is the solution in absence of the nonlinearities. Moreover,
\[
\widetilde{J}[h]=\frac{\lambda}{2}\left(\frac{\partial h}{\partial
x}\right)^{\!\!2}\widetilde{P}=\mbox{const}\;
\]
for the nonlinear case. A special feature of our discretization is
that  $\widetilde{P}(\mathbf{h})$ is a steady state solution for
all values of the $\lambda$. While the nonlinear coefficient is
present in $\widetilde{P}(\mathbf{h})$, it is absent in
$\widetilde{P}[h]$, because $\lim_{a\to 0}\sum_{i=1}^N a\,
U_i^{(\gamma)}=\int_0^L\left({\partial h}/{\partial
x}\right)^{\!2} dx$. This means that the coarse grained
approximation erase in $\widetilde{P}[h]$ our knowledge on the
discrete model and the nonlinearity. Again, for any
discretization, the coarse grained approximation preserves the
nonlinear dominant term of the steady state current, since
$\lim_{a\to 0}\widetilde{J}_i^{(\gamma)}=\widetilde{J}[h]$.

\emph{Ballistic deposition model.} The KPZ equation is tractable
by direct numerical integration when we specify its associate
spatial discretization. In order to verify this fact, we derive
the Langevin equation [Eq.~(\ref{eq:discrete-kpz})] for the
ballistic deposition (BD) model. The procedure used here is based
on regularizing the step functions included in the growth rules of
the microscopic model, in order to obtain the discrete Langevin
equation and, afterward coarse graining, the KPZ equation. Let us
first introduce the general treatment. The discrete interface
growth at an average rate $\tau$ and the interface height at an
$i$th-site increase in $h_i(t+\tau)-h_i(t)=a\,
\sum_{j=1}^m\,r_i^{(j)}$, where $r_i^{(j)}$ are the rules of the
deposition processes. Expanding $h_i(t+\tau)$ up to second order
in Taylor series around $\tau$, we obtain
$h_i(t+\tau)-h_i(t)\approx\tau\,dh_i/dt$. Thus, the height
evolution equation of the $i$th-site is given by the Langevin
equation
\begin{equation}\label{eq:Langevin}
\frac{dh_i}{dt}=K_i^{(1)}+\eta_i(t)\;,
\end{equation}
where the Gaussian thermal noise $\eta_i$ has zero mean and covariance
\begin{equation}\label{eq:cov-2-moment}
\langle\eta_i(t) \eta_j(t') \rangle = K_{ij}^{(2)}\;\delta (t -
t')\;.
\end{equation}
The first and second moments of the transition rate, in terms of growth rules, are given by
\begin{equation}\label{eq:moments}
K_i^{(1)}=\frac{a}{\tau}\,\sum_{j=1}^m\,r_i^{(j)}\;,\hspace{5ex}
K_{ij}^{(2)}=a\,\delta_{ij}\,K_1^{(1)}\;,
\end{equation}
respectively. In the BD model, a particle is released from
a randomly chosen lattice position $i$ above the interface,
located at a distance larger than the maximum height of the
interface. The incident particle follows a vertical straight
trajectory and sticks to the interface at time $t$. The height
in the column $i$ is increased by $\max[h_{i-1},h_i+1,h_{i+1}]$.
For this model the rules can be summarized as:
\begin{eqnarray}\label{eq:rules}
r^{(1)}_i&=&\Theta(H^i_{i+1})\; \Theta(H^i_{i-1})\;,\\
r^{(2)}_i&=&H^{i+1}_i\left[1-\Theta(H^i_{i+1})\right]\left[1-\Theta(H^{i-1}_{i+1})\right]\;,\nonumber\\
r^{(3)}_i&=& H^{i-1}_i\left[1-\Theta(H^i_{i-1})\right]\left[1-\Theta(H^{i+1}_{i-1})\right]\;,\nonumber\\
r^{(4)}_i&=&\textstyle\frac{1}{2}\; \delta(H^{i+1}_{i-1},0)\;
\left\{H^{i+1}_i\left[1-\Theta(H^i_{i+1})\right]\right.\nonumber\\
&&\hspace{14ex}+\left.H^{i-1}_i\left[1-\Theta(H^i_{i-1})\right]\right\}\;,\nonumber
\end{eqnarray}
where $\Theta(z)$ is the unit step function defined as
$\Theta(z)=1$ for $z\ge 0 $ and $\Theta(z)= 0$ for $z<0$, and
$\delta(z,0)=\Theta(z)+\Theta(-z)-1$ is the Kronecker delta. The
representation of the step function can be expanded as
$\Theta(z)\doteq\sum_{k=0}^{\infty} c_k z^k$ providing that $z$ is
smooth. In any discrete model there is in principle an infinite
number of nonlinearities, but at long wavelengths the higher order
derivatives can be neglected using scaling arguments, since one
expect affine interfaces over a long range of scales, and then one
is usually concerned with the form of the relevant terms. Thus,
keeping the expansion of the step function to first order in its
argument and replacing the expansion in Eq.~(\ref{eq:rules}) and
(\ref{eq:moments}) the first moment is
\begin{equation}\label{eq:1st-moment}
K_i^{(1)}=v_0+\nu\,L_i + \frac{\lambda}{2}\,N_i^{(\gamma)}\;,
\end{equation}
where
\begin{eqnarray}\label{eq:coef}
v_0 &=& c_0^2\,\frac{a}{\tau}\;,\nonumber\\
\nu &=& (1-c_0-2 c_0 c_1)\,\frac{a^2}{\tau}\;,\\
\lambda&=&2\,c_1\left(5-4
c_0-c_1\right)\;\frac{a}{\tau}\;,\nonumber\\
\gamma&=&\frac{1}{2}+\frac{1-2(c_0+c_1)}{2\,(3-2\,c_0)}\;.\nonumber
\end{eqnarray}
The average driving velocity $v_0$ can be subtracted in the
expression of the first moment given by Eq.~(\ref{eq:1st-moment}),
choosing adequately a moving reference frame. Retaining only the
constant term in Eq.~(\ref{eq:1st-moment}) we obtain
$K_{ij}^{(2)}\simeq a\,\delta_{ij}\,v_0=\epsilon\,\delta_{ij}/a$.
Replacing in Eq.~(\ref{eq:cov-2-moment}) we recover
Eq.~(\ref{eq:cov-discrete}) with noise intensity
$\epsilon=a^2\,v_0$. Notice that in order to integrate numerically
the continuous equation, we need a continuous representation of
the $\Theta$-function to numerically compute the coefficients
$c_0$ and $c_1$ related to the ones of the KPZ equation through
Eq.~(\ref{eq:coef}), such as the shifted hyperbolic tangent
representation \cite{Predota}. Our parameter of discretization is
related to the parameters of the microscopic model and the KPZ
coefficients [see Eq.~(\ref{eq:coef})]. Choosing $\gamma$, the KPZ
coefficients $\lambda$, $\nu$ and $\epsilon$ depends only of one
of the microscopic parameters, {\sl e.g.} with $\gamma=1/2$ the
parameters $c_0$ and $c_1$ are related as $c_0+c_1=1/2$.

In summary, we propose a finite difference discretization
criterion for the numerical integration of the KPZ growth equation
starting from the corresponding generalized discrete Langevin
equation. We show that, for any discretization scheme, exists a
steady state probability density function of the discrete
interface heights compatible with the continuous ones, after
coarse graining. Besides, we derived the KPZ coefficients of the
BD model as an example of our generalized discretization. Finally,
our results can be used as a new tool for the direct numerical
integration of growing continuous equations with nonlinear terms
due to lateral growth like the KPZ one.

Acknowledgements: We wish to thank M. Bellini for valuable
suggestions and L.A. Braunstein for a critical reading of
the manuscript. This work was supported by UNMdP and ANPCyT (PICT-O
2000/1-03-08974).

\end{document}